\documentclass[preprint,authoryear,12pt]{elsarticle}

\usepackage{amssymb}

\usepackage{graphicx}
\usepackage{url}
\usepackage{floatflt}
\usepackage[english]{babel}
\usepackage{latexsym}
\usepackage{algorithm}
\usepackage{algorithmic}
\usepackage{natbib}
\usepackage{amssymb,amsmath}
\usepackage[usenames]{color}
\usepackage[rgb,svgnames]{xcolor}
\usepackage{longtable}
\usepackage{array}
\usepackage{caption}
\usepackage{subcaption}
\usepackage{setspace}

\DeclareGraphicsExtensions{.eps, .jpg, .eps}

\newtheorem{asp}{Assumption}

\newcommand{\mv}{ {\mbox{v}} }

\newcommand{\SF}{ {\cal F} }

\newcommand{\ST}{ {\cal T} }

\newcommand{\SV}{ {\cal V} }

\newcommand{\SX}{ {\cal X} }

\newcommand{\bP}{ {\bf P} }

\newcommand{\bgamma}{ {\mbox{\boldmath $\gamma$}} }

\newcommand{\bnu}{ {\mbox{\boldmath $\nu$}} }

\newcommand{\bzr}{ {\bf 0} }

\usepackage{comment}

\begin{document}

\begin{frontmatter}



\vspace{-0.5cm}

\title{R-Score: Reputation-based Scoring of Research Groups}


\author[ufmg]{Sabir Ribas}\ead{sabir@dcc.ufmg.br}
\author[ufmg,google]{Berthier Ribeiro-Neto}\ead{berthier@dcc.ufmg.br}
\author[ufrj]{Edmundo de Souza e Silva}\ead{edmundo@land.ufrj.br}
\author[ufmg]{Nivio Ziviani}\ead{nivio@dcc.ufmg.br}

\address[ufmg]{Computer Science Dept., Universidade Federal de Minas Gerais, Brazil}
\address[ufrj]{Computer Science Dept., Universidade Federal do Rio de Janeiro, Brazil}
\address[google]{Google Engineering, Brazil}

\begin{abstract}

To manage the problem of having a higher demand for resources than availability of funds, 
research funding agencies usually rank the major research groups in their area of knowledge. 
This ranking relies on a careful analysis of the research groups in terms of their size, number of PhDs graduated, 
research results and their impact, among other variables. While research results are not the only variable to consider, 
they are frequently given special attention because of the notoriety they confer to the researchers and the programs they 
are affiliated with. 
In here we introduce a new metric for quantifying publication output, 
called {\em R-Score} for reputation-based score, which can be used in support to the ranking of 
research groups or programs. 
The novelty is that the metric depends solely on the listings of the publications of the members of 
a group, with no dependency on citation counts.
R-Score has some interesting properties: (a) it does not require access to the contents of 
published material, (b) it can be curated to produce highly accurate results, and 
(c) it can be naturally used to compare publication output of research groups (e.g., graduate programs) inside 
a same country, geographical area, or across the world. 
An experiment comparing the publication output of 25 CS graduate programs from Brazil suggests that R-Score 
can be quite useful for providing early insights into the publication patterns of the various research groups one wants to compare. 

\end{abstract}


\end{frontmatter}




\section*{Introduction}

\noindent Research financing in most countries is usually done by federally funded agencies. 
To allocate funds, these agencies usually rely on some form of ranking of the research groups (e.g., university departments or graduate programs) in all major areas of knowledge. This is normally accomplished through a time 
consuming and detailed process of evaluating and comparing publication records, number 
of students graduated, quality of the publications, international visibility, and focused surveys. 
In their evaluation processes, funding agencies usually rely on the publication records 
of professors and researchers as an important signal of productivity of a graduate program or research group 
operating inside a university department or research center. Thus, properly estimating and weighting publication 
records, with the purpose of comparing academic output of research groups, is a critical 
step of the evaluation process. 

However, comparing publication records 
is difficult to do in real life situations. To illustrate, 
one variable to consider is the {\em impact factor} ({\em IF}) of a publication venue, such as a major journal in a given 
area of knowledge. But computing IF requires access to the contents of all other publications that cite 
the articles published by that publication venue, which can only be done by building up very large publication 
record repositories such as those maintained by Google Scholar, Scopus, Web of Science (WoS), and Microsoft Academic 
Research. These repositories are costly to build, expensive to maintain, and their databases are not available unrestrictedly. 
Further, even the largest of these repositories store just a fraction of all publications in a 
given area of knowledge. As a result, any computation of IF is approximate and, more 
troublesome, the error incurred at the computation might lead to wrongful conclusions~\citep{bar-ilan2007}. 

\enlargethispage{-\baselineskip}

In this paper we focus on the problem of how to compare research groups, 
usually affiliated to a department in a university, based solely 
on the listings of publications of their faculty members. For this, we introduce a metric 
which we refer to as R-Score, for reputation-based score. Our idea is to compare the 
listings of publications of various research groups in a given area of knowledge, in light of the 
listings of publications of the top groups in that area. The intuition is that programs 
that publish frequently in the venues of preferences of the top programs in their area 
are more productive than programs that publish elsewhere. 

R-Score has three fundamental properties. First, it does not require access to the contents 
of the published material in a given area because it is not based on citation counts.  
All that is required is a list of all papers 
published by each program. Second, because of this it is simpler and can be curated quickly. 
Third, our experiments suggest that it can be used to provide early understanding of the 
research output of the programs one wants to compare. In this regard, R-Score can be seen 
as a useful metric to complement the complex and costly 
evaluation procedures run by funding agencies in modern countries. 

The remaining of this paper is organized as follows.
The next section covers related work.
Following, we discuss our reputation-based approach to compare the publication output of research groups. 
We then present our experiments and, at the end, our conclusions. 

\section*{Related Work}
\label{sec:relw}

\noindent Bibliometrics is a set of methods to quantitatively analyze both scientific and technological literature 
\citep{bellis2009bibliometrics,Mann:2006,Zhuang:2007,Yan:2007,Shaparenko:2007,He:2009}. Papers in the 
context of bibliometrics generally aim at improving existing metrics with text mining, improving other analysis by including bibliometric data, or proposing new metrics (as in this work).

Measuring the quality of publication venues is an important task in bibliometrics \citep{Zhuang:2007}. The most widely adopted 
approach to perform this task is to use Garfield's Impact Factor (IF) \citep{garfield1955}. Since its introduction,
IF has been criticized primarily due to its sole dependency on citation counts \citep{saha2003}. 
To address those issues, many alternatives have been proposed in the literature, 
such as the H-index \citep{hirsch2005}, download-based measures \citep{bollen2005toward}, and PageRank-like measures \citep{bollen2006,Yan:2007}, techniques which have been applied to rank computer and information science 
journals \citep{Katerattanakul2003,nerur2005assessing}. Also, several citation-based metrics have been 
proposed to rank documents retrieved from a digital library \citep{larsen2006using}, and to measure the quality of a small set of conferences and journals in the database field \citep{rahm2005citation}. 
In a recent study, \cite{Mann:2006} introduces topic modeling to further complement the citation-based bibliometric indicators, producing more fine-grained impact measures.
\cite{Yan:2007} propose two measures for ranking the impacts of academic venues -- an easy-to-implement 
seed-based measure that does not require citation analysis, and a realistic browsing-based 
measure that takes an article reader's behavior into account. 

To our knowledge, there has not been much work done in ranking academic research groups using bibliometrics. 
However, the problem is present in real life and is of high relevance. To exemplify, the 
United States National Research Council (NRC) regularly compiles survey-based rankings of US graduate programs. 
These rankings have two distinct statistical natures: R rankings and S rankings.
R Rankings use regression analysis of various survey results in which academics review the reputation of actual programs. S Rankings, on the other hand, are based on how various programs' characteristics measure against criteria which academics rate as key determinants of quality for such programs \citep{usna}.

Our work proposes a new bibliometric method for quantifying the publication output of 
research groups (e.g., graduate programs), a variable that often receives high weight in any 
ranking of research groups or academic programs. It depends neither on citations nor on document contents, relying solely on a set of the known top research groups in an area and on a list of all papers published by each group. 

\section*{R-Score--A Reputation-based Metric of Research Output} 
\label{sec:reputation-ranking}

Our main objective is to measure and compare the publication output of research groups 
that work in a given area of knowledge, based on the productivity 
of their members. In this paper, we focus our attention on graduate programs in Computer Science (CS) in particular, i.e., the 
research groups we consider are the faculty associated with CS graduate programs. Further, all 
graduate programs we refer to are run by a single CS Department of a university. Because of that, in the interest of 
objectivity and without loss of generality, in the remaining of this paper we use interchangeably {\em graduate programs and their 
faculty} or {\em departments and their professors}, instead of research groups and their members. 

The basic idea of our method is to compare the set
of programs to be ranked against that of a distinct set composed of the top peer programs in the world---with the caveat that these 
top programs need not to be ranked. 
Our intuition is that the top programs in any area confer authority to publication venues, 
which can then be used to assign weights to the venues. 
Given the venues are now weighted, we can rank the publication output of the programs we want to compare 
based on how they publish in these venues. 

\subsection*{Basic Considerations}

An important assumption of our proposal is that we can reasonably identify a set
of the most prestigious programs (that we refer to as the {\em top reference set}) 
in a given area and that we have access to the venues where the faculty members
of these programs publish.
Notice that distinct top reference sets might be considered, which will lead to distinct 
R-Score results. The important points to emphasize here are: (a) R-Score provides a comparison 
of publication output in view of a pre-selected set of top reference programs and 
(b) selecting a dozen or so top reference programs in a given area can be done much more 
objectively and with much smaller effort than weighting hundreds of publication venues in 
that area, or computing citation counts for papers in those venues. 

Regarding R-Score computation, it 
is important to emphasize that there is no need to access the contents of the
papers published by the faculty members of these programs and the citations to
their work.  
Also, there is no need to rank the programs in the {\em top reference set}.
What is required is simply to count the number of publications
by the faculty members of programs 
in the chosen {\em top reference set} at different venues.

\enlargethispage{\baselineskip}

Two issues that immediately arise is how to choose the programs in the {\em top reference set}
and how many programs we need to include in this set.
We argue that it is not difficult to identify a few dozens programs that would be
reasonable candidates to the {\em top reference set}. 
If we ask a set of senior professors in a given area of knowledge to name, for instance,
the top $10$ programs 
in the world in their area, we expect that a large fraction of all selected programs 
will be included in a large fraction of each of the professor's list.
For instance, if the area chosen is Computer Science, graduate programs run by 
departments at the Massachusetts Institute of Technology, Univerity of California at Berkeley, 
Carnegie Mellon University, and Stanford University will most likely be included 
in any {\em top reference set} list. 
In fact, a brief search in world wide academic rankings would support such claim.
We postpone the choice of the cardinality of the {\em top reference set} 
for future sections where we also study the sensitivity of the results we report
to the size of the {\em top reference set}.

In what follows, we describe the methodology we propose.
It is based on a {\em role model} concept, in which one should take as model
the most reputable individuals in our society.
We start by stating our main assumptions.

\begin{asp}
\label{asp:dep-fac}
The reputation of a graduate program is strongly influenced by the reputation of its faculty, 
which is largely dependent on their publication track records. 
\end{asp}
As a consequence of this assumption, the graduate programs in the {\em top reference set} employ the most prestigious faculty.
The more prestigious is a program the better chances it has to attract the most prominent Ph.D. graduates and
senior renowned scientists. 

\begin{asp}
\label{asp:fac-ven}
A researcher, or member of a graduate program, conveys reputation to a venue proportionally to its own reputation.
\end{asp}
This assumption is a consequence of what we call the role-model effect.
Reputable scientists usually choose prestigious venues to publish at and, as
such, the reputation of a venue is positively correlated with the reputation of the individuals that publish in
that venue.
As more prestigious researchers of an area choose a venue to publish their work, the venue becomes increasingly 
known by peer researchers and, as a consequence, attracts even more distinguished researchers and
young scientists, building up its reputation.   

\begin{asp}
\label{asp:ven-fac}
The reputation of a faculty member is positively correlated with the reputation of the venues in which he/she
publishes.
\end{asp}
One of the most used metrics to promote a faculty member in any reputable department or graduate program 
is the number of papers in prestigious venues where the faculty under consideration publishes. 
Clearly, if a given scientist has a reasonable number of papers in the most prestigious venues
then it is reasonable to assume he/she is a prestigious scientist.

The three assumptions above form the cornerstone of our reputation-based ranking model.
Our model is inspired by the {\em eigenvalue centrality} metrics for complex networks
\citep{Newman-book}, which operate quite similarly to Page Rank \citep{Meyer-pagerank06}.
This metric is obtained by assuming
that the relative importance (or reputation)
of a node in a network is proportional to the importance of the nodes that point to it.
In addition, a given node distributes its importance uniformly among the nodes it points to.
We map these ideas to obtain a ranking model for a set of venues.

\subsection*{Notation and Publication Counts}

Before developing the model, we introduce some notation.
We use $\omega$, $i$ and $j$ as indexes for graduate programs, their faculty, and the venues where they publish, respectively.
The graduate programs used as reputation sources are referred to jointly as the {\em top reference set}. 
Consider a chosen set $\ST$ of top reference programs, and let $T$ be its cardinality.
In addition, let $\SF_\omega$ be the set of faculty members of program $\omega$, 
$f_{i\omega}$ be the $i^{th}$ faculty member of program $\omega$, and 
$F_\omega = | \SF_w |$ be the total number of faculty members in $\omega \in \ST$.
Let $\SV$ be the set of all venues $\mv_j$ where the faculty in $\ST$ publish, and $V$ the total number of
venues in the set $\SV$. 
Faculty members of program $\omega$ publish in subset $\SV_\omega \subseteq \SV$ with cardinality 
$V_\omega = | \SV_\omega |$.
Denote by $\gamma_{i\omega}$ the reputation of faculty $i$ of program $\omega \in \ST$,
and $\nu_j$ the reputation of venue $\mv_j \in \SV$.

\enlargethispage{2\baselineskip}

Let us define a function $N$ that counts the papers published by faculty, the papers published by whole programs, and the papers published at venues. 
In this regard, let $N(f_{i\omega}, \mv_j)$ be the total number of papers published by faculty member $i$
of program $\omega$ in venue $\mv_j$ during a given period of time, weighted by the number of co-authors that are also
faculty at program $\omega$. Also, let $a_{k (i\omega) j}$ be the number of faculty at program $\omega$ that co-authors of the $k^{th}$ paper 
published in venue $\mv_j$ by faculty member $f_{i\omega}$. 
If a paper has a single author 
from program $\omega$, then $a_{k (i\omega) j} = 1$. It follows that:
\begin{equation}
\label{eq:n-fwi-vj}
N(f_{i\omega}, \mbox{v}_j) = \sum_{k} \frac{1}{a_{k (i\omega) j}}
\end{equation} 
The motivation here is that if a given paper is co-authored by $h$ authors belonging to the same program, that
paper counts as $1/h$ to the number of papers published by each of its co-authors in that program.
This leads immediately to:
\begin{equation}
\label{eq:psi}
N(\omega, \mbox{v}_j) = \sum_{i=1}^{F_w} N(f_{i\omega}, \mv_j)
\end{equation} 
which counts the total number of {\em distinct papers} 
published by program $\omega$ in venue $\mv_j$.

Complementing the definitions of function $N$, let $N(\mbox{v}_j)$ and $N(w)$ be the total number of papers published in 
venue $\mv_j$ and
the total number of publications of program $\omega$ during the observation period, respectively.
That is:

\begin{eqnarray*}
N(w) & = & \sum_{j=1}^{V} N(\omega, \mbox{v}_j) \\
N(\mbox{v}_j) & = & \sum_{w=1}^{T} N(\omega, \mbox{v}_j) 
\end{eqnarray*}
Note that, in the sum to obtain $N(\mbox{v}_j)$, we are considering that joint publications from faculty
at different programs count more than once to the sum. 
An alternate definition to $N(\mbox{v}_j)$ is to count joint publications only once in $N(\mbox{v}_j)$
and then divide the relative importance of these publications uniformly among co-authors in different programs.
However, we find the first definition more appropriate to our studies,
both to emphasize the contribution to a venue from distinct programs
and because this is usually the way publications are counted by programs.
Nevertheless, our model supports both definitions.
Table \ref{t:notation} summarizes the above notation and definitions. 
\begin{center}
\begin{footnotesize}
\begin{longtable}{|c|p{10.0cm}|}
\hline
$\ST,\;\; (T)$              &set of top reference programs (cardinality of $\ST$) \\
$\omega$                    &a graduate program in the top reference set \\
$\SV,\;\; (V)$              &set of venues where the researchers in $\ST$ publish (cardinality of $\SV$) \\
$\SV_\omega,\;\; (V_{\omega})$   &set of venues where the researchers of program $\omega$ publish (cardinality of $\SV_\omega$) \\
$\mbox{v}_j$                       &the $j^{th}$ venue where faculty of a top reference set program publishes at \\
$f_{i\omega}$            &the $i^{th}$ faculty member of program $\omega$ \\
$a_{k (i\omega) j}$        &number of program $\omega$ co-authors of the $k^{th}$ paper published by $f_{i\omega}$ in venue $\mv_j$ \\ 
$N(f_{i\omega}, \mbox{v}_j)$              &total number of papers published by faculty member $i$ of program $\omega$ in venue $\mv_j$,
                             weighted by the number of faculty co-authors in the same program \\
$N'(f_{i\omega}, \mbox{v}_j)$              &total number of papers published by faculty member $i$ of program $\omega$ in venue $\mv_j$ \\
$N(\omega, \mbox{v}_j)$                 &total number of distinct papers published by program $\omega$ in venue $\mv_j$ \\
$N(\mbox{v}_j)$                       &total number of papers published in venue $\mv_j$ \\
$N(w)$                       &total number of publications of program $\omega$ \\
$\gamma_{i\omega}$         &reputation of faculty $i$ of program $\omega \in \ST$ \\
$\gamma_{\omega}$         &reputation of program $\omega \in \ST$ \\
$\nu_j$                     &reputation of venue $\mv_j \in \SV$ \\ 
\hline
\caption{Notation.}
\label{t:notation}
\end{longtable}
\end{footnotesize}
\end{center}

It is convenient, at this point, to discuss a small example to illustrate the notation
before proceeding with the model development. 
%
%

\subsection*{Example}

\begin{figure}[htb]
   \centering
   \includegraphics[scale=0.68]{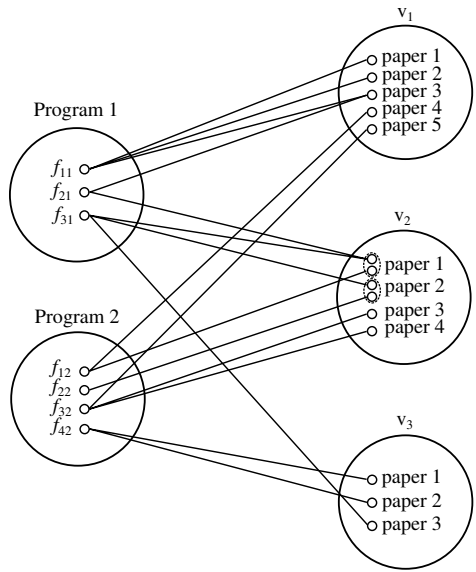} 
   \caption{A small example.}
   \label{fig:ex1} 
\end{figure}

Figure \ref{fig:ex1} shows an example with two programs in the {\em top reference set} and three venues in set 
$\SV$. 
The circles on the left represent the programs in the {\em top reference set} and the dots inside these circles are the faculty members.
Venues are represented by the circles on the right and the dots inside them indicate the different papers published in each venue. 
Using our notation, $\ST = \{\mbox{Program $1$},\; \mbox{Program $2$} \} $, $T=2$, $\SV = \{ \mbox{v}_1, \mbox{v}_2, \mbox{v}_3 \}$ and $V = 3$. 
Also, in this particular case, $\SV_1 = \SV_2 = \{ \mbox{v}_1, \mbox{v}_2, \mbox{v}_3 \}$ and $V_1 = V_2 = 3$.
From Figure \ref{fig:ex1}, faculty $f_{11}$ from Program 1 published $3$ papers in venue $\mbox{v}_1$ and
her last paper is co-authored with faculty $f_{21}$ from the same program. 
From our notation, 
\[
\begin{array}{ccc}
N(f_{11}, \mbox{v}_1) = 1 + 1 + \frac{1}{2} = 2.5; &
N(f_{11}, \mbox{v}_2) = 0; &
N(f_{11}, \mbox{v}_3) = 0 \\[0.2cm]
\end{array}
\]
Faculty $f_{31}$ of Program $1$ also published $3$ papers: one co-authored
with another colleague of the same program, one co-authored with a colleague from Program $2$, 
and a third paper co-authored with no other faculty. 
In our notation, 
\[
\begin{array}{ccc}
N(f_{31}, \mbox{v}_1) = 0; &
N(f_{31}, \mbox{v}_2) = 1.5; &
N(f_{31}, \mbox{v}_3) = 1 \\[0.2cm]
\end{array}
\]
Notice that, in the representation of venue $\mbox{v}_2$, papers numbered $1$ and $2$ are each represented by two dots, since each of them has co-authors
from different programs. This is done for illustrative purposes and in accordance with our definitions.

From the explanation above and using Figure \ref{fig:ex1}, it is easy to infer 
the publications and co-authorship of faculty belonging to Program 2. 
Using Equation (\ref{eq:psi}) the number of distinct papers published by a program in a venue is:
\[
\begin{array}{ccc}
N(1, \mbox{v}_1) = 3; & N(1, \mbox{v}_2) = 2; & N(1, \mbox{v}_3) = 1 \\
N(2, \mbox{v}_1) = 2; & N(2, \mbox{v}_2) = 4; & N(2, \mbox{v}_3) = 2 \\[0.2cm]
\end{array}
\]
Since, by the definition of $N(\mbox{v}_j)$, the first and second papers in venue $\mbox{v}_2$ count as one unit for each program, 
we can count the number of papers in each of the three venues as follows:
\[
\begin{array}{ccc}
N(\mbox{v}_1) = 5; & N(\mbox{v}_2) = 6; & N(\mbox{v}_3) = 3 \\[0.2cm]
\end{array}
\]
Furthermore, we count the number of papers published by each of our two programs as follows:
\[
\begin{array}{cc}
N(1) = 6; & N(2) = 8 \\[0.2cm]
\end{array}
\]

\subsection*{A Markov Model of Reputation}

We continue with the development of our model of reputation. 
From Assumption (\ref{asp:ven-fac}) and further assuming that the reputation of a faculty member
is only given by the reputation of the venues he/she publishes, we have:
\begin{equation}
\label{eq:rep-fac}
\gamma_{i\omega} = \sum_{j=1}^V \frac {N(f_{i\omega}, \mbox{v}_j)} {N(\mbox{v}_j)} \times \nu_j
\end{equation}

Equation (\ref{eq:rep-fac}) immediately follows from 
{\em eigenvalue centrality} concepts from which the reputation of
a venue is uniformly distributed among the total number of papers in that venue published by
all researchers in the {\em top reference set} programs. 
Note that we assume that the reputation of a given faculty member $i$
is given solely by the reputation of the venues he/she publishes in proportion to the number of papers
published by $i$ at each of the venues.
In Equation (\ref{eq:rep-fac}), we could alternatively use the other definition for $N(\mbox{v}_j)$ referred above,
that is, counting only the number of distinct papers in $\mv_j$.
In this case, a similar 
result would be obtained for $\gamma_{i\omega}$ provided that,
for each paper jointly published by members of different programs, 
the portion of the venue's reputation due to that paper transferred to
faculty $i$ would be divided equally among the faculty co-authors in different programs. 

From Assumption (\ref{asp:dep-fac}) we consider that the reputation of a program $\omega$ is directly
proportional to the reputation of its faculty members.
Therefore, summing the individual reputation of its faculty given in Equation (\ref{eq:rep-fac}) we have:
\begin{eqnarray}
\gamma_w & = & \sum_{i=1}^{F_w} \gamma_{i\omega} \nonumber \\ 
         & = & \sum_{j=1}^V \nu_j \times \alpha_{wj} \label{eq:rep-dep}
\end{eqnarray}
where 
\begin{equation}
\label{eq:alpha}
\alpha_{wj} = \frac {N(\omega, \mbox{v}_j)} {N(\mbox{v}_j)}
\end{equation}
is the fraction of publications of venue $\mv_j$ that are from program $\omega$. 

Equation (\ref{eq:rep-dep}) obtains the reputation of a program in the {\em top reference set} $\ST$ from the reputation
of the set $\SV$ of venues.
We now employ Assumption (\ref{asp:fac-ven}) to obtain the reputation of all venues in set $\SV$.
For this goal, we resort to the {\em eigenvalue centrality} concepts again and assume that the reputation of a program 
is uniformly distributed to all the papers published by its faculty members.
We have: 
\begin{equation}
\label{eq:rep-ven}
\nu_j = \sum_{w=1}^T \gamma_w \times \beta_{wj} 
\end{equation}
where
\begin{equation}
\label{eq:beta}
\beta_{wj} = \frac {N(\omega, \mbox{v}_j)} {N(w)}
\end{equation}
is the fraction of publications of program $\omega$ that are from venue $\mv_j$.

Let $\bP$ be $(T+V) \times (T+V)$ square matrix such that element $p_{mn} = 0$ if 
either $m, n \leq T$ or $m, n \geq T$.
In addition, $p_{mn} = \beta_{m,n-T}$ for $m \leq T, n > T$ and   
$p_{mn} = \alpha_{m-T,n}$ for $m > T, n \leq T$.
Note that, since $\sum_{w=1}^{T} \alpha_{wj} = 1$ for all $ 1 \leq j \le V$ and 
$\sum_{j=1}^{V} \beta_{wj} = 1$ for all $1 \leq w \leq T$ then
$\bP$ defines a Markov chain.
In addition, the Markov chain is periodic and has the following structure:
\[
\bP =
\left[
\begin{array}{c c c c | c c c c}
0            &0             &\ldots    &0           &\beta_{11} &\beta_{12}   &\ldots   &\beta_{1V} \\
\vdots       &              &\ddots    &\vdots      &\vdots     &\vdots       &\ddots   &\vdots    \\
0            &0             &\ldots    &0           &\beta_{w1} &\beta_{w2}   &\ldots   &\beta_{wV} \\
\vdots       &              &\ddots    &\vdots      &\vdots     &\vdots       &\ddots   &\vdots    \\
0            &0             &\ldots    &0           &\beta_{T1} &\beta_{T2}   &\ldots   &\beta_{TV} \\
\hline 
\alpha_{11}  &\alpha_{21}   &\ldots    &\alpha_{T1} &0           &0           &\ldots   &0  \\
\vdots       &\ddots        &\vdots    &\vdots      &\vdots      &            &\ddots   &\vdots \\
\alpha_{1j}  &\alpha_{2j}   &\ldots    &\alpha_{Tj} &0           &0           &\ldots   &0      \\
\vdots       &\ddots        &\vdots    &\vdots      &\vdots      &            &\ddots   &\vdots \\
\alpha_{1V}  &\alpha_{2V}   &\ldots    &\alpha_{TV} &0           &0           &\ldots   &0 
\end{array}
\right]
= 
\left[
\begin{array}{c | c}
\bzr      &\bP_{12} \\
\hline
\bP_{21}  &\bzr    \\
\end{array}
\right]
\]
\noindent 
From decomposition theory, see \cite{meyer89}, we can obtain values for 
ranking the {\em top reference set} programs by solving:
\begin{equation}
\label{eq:top-rank}
\bgamma = \bgamma \bP^\prime
\end{equation}
where $\bP^\prime = \bP_{12} \times \bP_{21}$ is a stochastic matrix
and $\bgamma = \langle \gamma_1, \ldots, \gamma_T \rangle$.
Note that matrix $\bP^\prime$ has dimension $T \times T$ only and can be easily solved by
standard Markov chain techniques such as the GTH algorithm \citep{GTH85}.
Then, from Equation (\ref{eq:rep-dep}) we obtain the reputation of all venues where the top ranked programs publish.
\begin{equation}
\label{eq:venue-rank}
\bnu = \bgamma \times \bP_{12}
\end{equation}

\enlargethispage{-\baselineskip}

Once we obtain vector $\bnu$, which yields the ranking of the venues according to the programs in our
selected {\em top reference set} $\ST$, we can easily rank other programs not in $\ST$.
In our methodology, programs whose professors publish in the venues of choice
of the top programs' faculty
will be ranked higher than programs whose professors publish in other venues.
We emphasize that the key feature of our methodology is that, contrary to citation-based 
ranking functions, no access to the contents of the publications is required.

Returning to our small example of Figure~\ref{fig:ex1}, we can calculate the elements of matrix $\bP$ using
Equations (\ref{eq:alpha}) and (\ref{eq:beta}), as follows. 
%
%

\subsection*{Example (cont.)}

\begin{figure}[htb]
   \centering
   \includegraphics[scale=0.68]{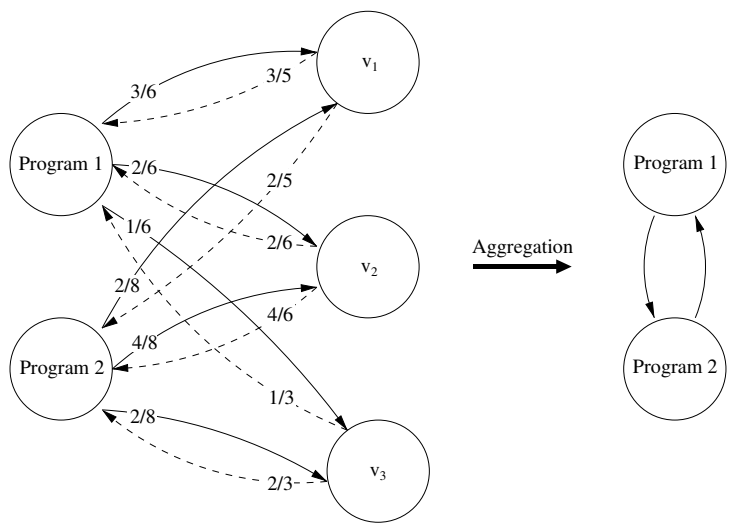}
   \caption{Markov chain for the small example of Figure~\ref{fig:ex1}.}
   \label{fig:ex1-MC}
\end{figure}

Figure~\ref{fig:ex1-MC} illustrates the Markov chain associated with the example in Figure~\ref{fig:ex1}. 
We have:
\[
\bP =
\left[
\begin{array}{c c | c c c }
0            &0             &3/6       &2/6         &1/6    \\
0            &0             &2/8       &4/8         &2/8    \\
\hline
3/5          &2/5           &0         &0           &0       \\
2/6          &4/6           &0         &0           &0       \\
1/3          &2/3           &0         &0           &0       \\
\end{array}
\right]
\]
This stochastic matrix corresponds to the Markov chain displayed in Figure \ref{fig:ex1-MC}, which
can be immediately aggregated to a two-state Markov chain, as shown in the figure, yielding
\[
\bP^\prime = 
\left[
\begin{array}{c c }
0.467   &0.533 \\
0.400   &0.600
\end{array}
\right]
\]
which is the stochastic matrix we use in the solution of Equation (\ref{eq:top-rank}).
(Recall that the dimension of $\bP^\prime$ is 
$T \times T$ and, as such, much smaller than that of $\bP$ for real size problems.)
Solving Equation (\ref{eq:top-rank}), applying Equation (\ref{eq:venue-rank})
and then normalizing the result such that the venue with the largest ranking is set to one,
we obtain the ranking for the three venues: $\bnu = \langle 0.83, 1.0, 0.5 \rangle$.
That is, venue $\mbox{v}_2$ has the highest rank, followed by $\mbox{v}_1$, and then by $\mbox{v}_3$. 
We remark that the individual values give the {\em relative importance} of each venue with respect to $\mbox{v}_2$.

\subsection*{R-Score Formula}

Let $x$ be a program in the set $\SX$ (distinct and disjoint from $\ST$) composed of those programs we want to rank.
We first obtain $N(x,\mv_j)$, that is, the total number of publications from program $x$ 
in venue $\mv_j \in \SV$.
Then we define R-Score as the rank $\gamma_{x}$ of program $x \in \SX$, normalized by the highest
rank in $\SX$:
%
\begin{equation}
\label{eq:d-rank}
\mbox{R-Score}(x \in \SX) = \frac{\gamma_{x}}{\max_{j \in \SX} \{ \gamma_j \} } 
\end{equation}
where, in this case, we define 
\begin{equation}
\label{eq:gamma_chi}
\gamma_{x} = \sum_{j=1}^V \nu_j \times N(x,\mv_j)
\end{equation}

As we discuss in our experiments, R-Score performs quite well relatively
to a much more costly citation-based metric, such as H-Index, for the purpose of comparing 
the publication records of distinct graduate programs. 

\section*{Evaluating CS Graduate Programs in Brazil}
\label{sec:comparing-cs-programs} 

In this paper we examine the problem of comparing Computer Science graduate programs in Brazil. 
Our interest is due to the many years of effort invested in complex evaluation procedures by the 
Brazilian research financing agencies and by our own familiarity with the issue. We focus on the 
top 25 Computer Science (CS) graduate programs in Brazil, according to the ranking of the CAPES 
funding agency (details below). 

We focus on comparing the ranking provided by CAPES with one generated directly by the application 
of R-Score. In this case, we consider as {\em reference set} 
the top 10 CS graduate programs in the US, as determined by the R Rankings of the National Research 
Council (NRC), $5^{th}$ 
percentile, for the year of 2010~\citep{nrc}, as shown in Table~\ref{t:topCSdept}. Before discussing 
our results, let us overview the evaluation procedure adopted by CAPES. 

\begin{table}[ht]
\caption{Top 10 CS graduate programs in the US in 2010, according to the R Rankings of NRC, $5^{th}$ percentile.}
\centering
\begin{tabular}{|l|}
\hline
1. Stanford University \\ 
2. University of California, Berkeley \\ 
3. Massachusetts Institute of Technology \\
4. Carnegie Mellon University \\
5. University of Illinois at Urbana-Champaign \\
6. Princeton University \\
7. Cornell University \\
8. University of California, Santa Barbara \\
9. University of North Carolina at Chapel Hill \\
10. University of California, Los Angeles \\
\hline
\end{tabular}
\label{t:topCSdept}
\end{table}

\subsection*{The CAPES Ranking of Brazilian Graduate Programs}
\label{sec:capes-rankings}

A well structured effort to evaluate graduate programs is the CAPES 
ranking in Brazil, which has been evaluating and comparing graduate 
programs since 1977, on a triennial basis~\citep{laender2008}. 
The process conducted by CAPES takes into account various quantitative and qualitative parameters such as coverage of 
courses' contents, curriculum vitae of professors, international reputation, 
number of master thesis and PhD dissertations concluded during the evaluation period, 
and publication records. Of these, one 
of the key parameters is the publication record of professors both in volume and in quality. 
This is also 
the most difficult parameter to estimate 
and any serious ranking of graduate programs must take into account a 
metric to quantify publication records. 

One simplistic approach to quantify publication records would be to count the number of papers 
published by each program or research group. This obviously does not work because it does not 
assign higher weights to high quality 
venues, exactly those that have highest impact, are more selective, and tend to publish fewer papers. Citation 
counts, while costly and difficult to compute, provide one possible answer to this problem. A 
slightly different approach has been adopted by CAPES, as follows. 

To compare publication 
records, CAPES classifies the publication venues in each area as A1, A2, B1, B2, B3, B4 and B5. Journals are ranked in each stratum based mostly on citation indexes (such as JCR or H-Index). For Computer Science, conferences are ranked based on the H-Index obtained from Google Scholar, and existing conference ranking such as the Computing Research and Education Association of Australia. This is a time-consuming 
and demanding task executed by committees formed by university professors. 

\enlargethispage{-2\baselineskip}

CAPES committees run a thorough comparative analysis 
of the publication records of the professors in each major graduate program in Brazil to establish a ranking of the 
programs. Each graduate program receives a grading in a scale of 3-7, where 7 is the highest ranking. To illustrate, there are more than four hundred CS departments in Brazil\footnotemark. 
Of these, a little over 50 CS departments have graduate programs with a ranking of 3 or higher. The following 
25 programs are the ones ranked 4 or higher (in parenthesis, we show the corresponding acronym and the 
current number of active professors):
\begin{itemize}
\item {\em Rank $7$:} Pontif\'{i}cia Universidade Cat\'{o}lica do Rio de Janeiro (PUC-Rio, 31), Universidade Federal de Minas Gerais (UFMG, 30), Universidade Federal do Rio de Janeiro (UFRJ, 36)
\item {\em Rank $6$:} Universidade Federal de Pernambuco (UFPE, 46), Universidade Federal do Rio Grande do Sul (UFRGS, 49), Universidade Estadual de Campinas (UNICAMP, 41), Universidade de S\~ao Paulo, S\~ao Carlos (USP-SC, 64)
\item {\em Rank $5$:} Universidade Federal Fluminense (UFF, 32), Universidade de S\~ao Paulo, S\~ao Paulo (USP-SP)
\item {\em Rank $4$:} Pontif\'{i}cia Universidade Cat\'{o}lica do Paran\'{a} (PUC-PR, 20), Pontif\'{i}cia Universidade Cat\'{o}lica do Rio Grande do Sul (PUC-RS, 22), Universidade Federal do Amazonas (UFAM, 19), Universidade Federal da Bahia (UFBA, 11), Universidade Federal do Cear\'a (UFC, 17), Universidade Federal de Campina Grande (UFCG, 21), Universidade Federal do Espirito Santo (UFES, 22), Universidade Federal do Mato Grosso do Sul (UFMS, 15), Universidade Federal do Paran\'a (UFPR, 25), Universidade Federal do Rio Grande do Norte (UFRN, 22), Universidade Federal de Santa Catarina (UFSC, 31), Universidade Federal de S\~ao Carlos (UFSCar, 25), Universidade Federal de Uberl\^andia (UFU, 17), Universidade de Brasilia (UnB, 13), Universidade de Fortaleza (Unifor, 17), Universidade do Vale do Rio dos Sinos (UNISINOS, 11)
\end{itemize}

\noindent CAPES considers that graduate programs with a rank of 6 and 7 are elite programs, that they are comparable to good programs abroad, and that they are those that shall receive funding from special programs. 

In here, we compare the ranking provided by CAPES for these 
top 25 graduate programs with a reputation-based ranking based solely on R-Score. 
We should keep in mind that while R-Score is based just on reputation and publication output, 
the official CAPES ranking considers many other variables such as 
the size of the program, the number of PhDs graduated, the history of publications and their impact.

\section*{Experiments}
\label{sec:experiments}

In our experiments we focused on comparing programs from the area of Computer Science, as discussed in 
previous sections. Our objective is not to propose an exact rank for each of these institutions, but 
instead to see how a ranking of academic programs based just on publication output 
compares with the much more complex ranking computed by CAPES.
For this reason, in here we
assign an alternative label to each program, composed of a letter and a number. Each letter maps to a CAPES rank (`A' mapping to rank 7, `B' mapping to rank 6, `C' mapping to rank 5, `D' mapping to rank 4), but there's no order whatsoever among the different numbers with the same letter. For example, programs with CAPES rank 6 are labeled as B1, B2, B3, and B4, without any implicit order established among them.

To determine the list of professors of each program, we extracted from the programs' official homepages the list of names of the faculty members.
To determine the publications by each professor of a program, we relied on the 
DBLP - Digital Bibliography \& Library Project\footnotemark \citep{ley@spire02} repository. 
While it is not exhaustive, it is extensive and covers all the major publication venues in the area. 

\subsection*{Ranking CS Graduate Programs Using R-Score}
\label{sec:r-score-ranking}

Using the discussed Markovian model, we computed R-Score values for the 25 CS programs in Brazil we want to compare. 
The results are presented in Table~\ref{t:r-score-total}.

\begin{table}[ht]
\caption{R-Score values for the 25 CS programs in Brazil, normalized.}
\begin{minipage}[b]{0.48\linewidth}
\centering
\begin{tabular}{|l|c|}
\hline
R-Score & CAPES Rank \\ \hline\hline
1.000000	&	B1 (6)	\\	
0.857428	&	A1 (7)	\\	
0.844877	&	B2 (6)	\\	
0.791402	&	A2 (7)	\\	
0.747391	&	A3 (7)	\\	
0.566406	&	C1 (5)	\\	
0.508885	&	B4 (6)	\\	
0.450852	&	B3 (6)	\\	
0.347899	&	C2 (5)	\\	
0.233254	&	D2 (4)	\\	
0.206170	&	D1 (4)	\\	
0.184552	&	D5 (4)	\\	
0.183148	&	D3 (4)	\\	
\hline
\end{tabular}
\end{minipage}
\begin{minipage}[b]{0.48\linewidth}
\centering
\begin{tabular}{|l|c|}
\hline
R-Score & CAPES Rank \\ \hline\hline
0.171225	&	D6 (4)	\\	
0.153983	&	D8 (4)	\\	
0.147344	&	D10 (4)	\\	
0.134397	&	D4 (4)	\\	
0.077107	&	D12 (4)	\\	
0.075324	&	D7 (4)	\\	
0.070584	&	D9 (4)	\\	
0.066620	&	D11 (4)	\\	
0.052814	&	D13 (4)	\\	
0.049478	&	D14 (4)	\\	
0.047381	&	D16 (4)	\\	
0.042542	&	D15 (4)	\\	
\hline
\end{tabular}
\end{minipage}
\label{t:r-score-total}
\end{table}

We observe that, despite its conceptual simplicity, R-Score yields a ranking of programs (by publication 
output) that matches 
quite well the ranking done by CAPES through an incomparably more sophisticated and time-consuming process. 
More important, R-Score provides a clear separation between the more productive programs, ranked 5 or higher, and the programs ranked at level 4. 

We further notice
that the distinct programs have rather varying numbers of faculty working in their graduate programs. 
To attenuate the effect of the number of faculty members, we also present a ranking considering the R-Score values divided by the number of faculty members in each program, as shown in Table~\ref{t:r-score-prof}.

\begin{table}[ht]
\caption{R-Score divided by the number of professors for the 25 CS programs in Brazil, normalized.}
\begin{minipage}[b]{0.48\linewidth}
\centering
\begin{tabular}{|l|c|}
\hline
R-Score/Prof & CAPES \\
Normalized & Rank \\ \hline\hline
1.000000	&	A1 (7)	\\	
0.894152	&	A2 (7)	\\	
0.730314	&	A3 (7)	\\	
0.723093	&	B1 (6)	\\	
0.710377	&	B2 (6)	\\	
0.487576	&	C1 (5)	\\	
0.405747	&	C2 (5)	\\	
0.397658	&	D8 (4)	\\	
0.391458	&	B4 (6)	\\	
0.324354	&	D2 (4)	\\	
0.324087	&	D1 (4)	\\	
0.294789	&	D6 (4)	\\	
0.290105	&	D5 (4)	\\	
\hline
\end{tabular}
\end{minipage}
\begin{minipage}[b]{0.48\linewidth}
\centering
\begin{tabular}{|l|c|}
\hline
R-Score/Prof & CAPES \\
Normalized & Rank \\ \hline\hline
0.254679	&	D3 (4)	\\	
0.246976	&	B3 (6)	\\	
0.231617	&	D10 (4)	\\	
0.212661	&	D9 (4)	\\	
0.142914	&	D4 (4)	\\	
0.136165	&	D7 (4)	\\	
0.126717	&	D12 (4)	\\	
0.122360	&	D16 (4)	\\	
0.120432	&	D11 (4)	\\	
0.106082	&	D13 (4)	\\	
0.096131	&	D15 (4)	\\	
0.066254	&	D14 (4)	\\	
\hline
\end{tabular}
\end{minipage}
\label{t:r-score-prof}
\end{table}

\enlargethispage{\baselineskip}

We notice that now the three programs that CAPES ranks as 7 appear on top. Further, 
programs with a large number of faculty members, such as B3 and B4, are penalized in the R-Score ranking, something that seems not to be the case with the CAPES ranking. 
That is, a comparison of Tables~\ref{t:r-score-total} and~\ref{t:r-score-prof} suggests that CAPES 
places a higher weight on the accumulated history of publication of a departament (over time) than on its present rate of publication (which is somewhat expected, if one wants to be conservative). 
In addition to publications, CAPES gives reasonable weight to the number of PhD and master students graduated in the
evaluation period.  Relatively young programs, as compared with traditional programs, have a smaller number of faculty members.
Therefore, although productive small young programs may have a high R-Score (normalized by the number of faculty members) they
do not achieve the necessary threshold in terms of number of students graduated to achieve the top grading levels from CAPES.

It is importat to emphasize that R-Score provides a metric of publication output to be used in support of 
a ranking of research groups. That is, R-Score was not conceived as a integral and complete ranking method of 
research groups by itself. Despite that, whenever the actual ranking of groups or programs, as done by major 
funding agencies, is heavily influenced by publication records, as is the case of the Brazilian agency CAPES, 
R-Score becomes quite an accurate predictor of the final ranking, as we have illustrated here. 

\subsection*{Stability of R-Score}

The R-Score method we proposed here uses a set of top programs to rank other programs. 
In previous sections, we have shown the effectiveness of this approach using the top 10 Computer Science faculties 
in the world to rank the publication output of the top 25 CS programs in Brazil. But, a natural and important question is what happens 
if we change the size of the top set, instead of using exactly 10 programs as the top reference set. That is, there 
is a question about how {\em stable} R-Score is. 

%

To measure the impact of using different sizes for the top reference set, we perform the following experiment.
Let $\mathit{Top}(x)=\{\mathit{Top}_1,\mathit{Top}_2,...,\mathit{Top}_x\}$ be a set composed by the top $x$ faculties of a given area of knowledge. For example, according to Table \ref{t:topCSdept}, $\mathit{Top}(3)=\{$Stanford University,\; Princeton University,\; Massachusetts Institute of Technology$\}$ in the context of Computer Science. Also, let $R_{\mathit{Top}}(x)$ be the ranking produced considering the set $\mathit{Top}(x)$ as source of reputation.
In the first step of our experiment, we produced ten rankings considering different top reference sets. 
Specifically, we generated $R_{\mathit{Top}}(i)$, $\forall i \in \{1,\cdots,10\}$. Next, we compared these rankings using Spearman's rank correlation coefficient \citep{spearman}. Table \ref{t:stability} presents the results. 

\begin{table}[h!t]
\caption{Comparison between rankings produced using different sizes of the \textit{top reference set}, according to the Spearman's rank correlation coefficient.}
\centering
\begin{tabular}{|l|r|}
\hline
Comparison	&Agreement\\\hline
$R_{\mathit{Top}}(1)$ versus $R_{\mathit{Top}}(2)$	&99.38\%\\
$R_{\mathit{Top}}(2)$ versus $R_{\mathit{Top}}(3)$	&99.54\%\\
$R_{\mathit{Top}}(3)$ versus $R_{\mathit{Top}}(4)$	&99.38\%\\
$R_{\mathit{Top}}(4)$ versus $R_{\mathit{Top}}(5)$	&99.23\%\\
$R_{\mathit{Top}}(5)$ versus $R_{\mathit{Top}}(6)$	&100.00\%\\
$R_{\mathit{Top}}(6)$ versus $R_{\mathit{Top}}(7)$	&99.54\%\\
$R_{\mathit{Top}}(7)$ versus $R_{\mathit{Top}}(8)$	&99.69\%\\
$R_{\mathit{Top}}(8)$ versus $R_{\mathit{Top}}(9)$	&100.00\%\\
$R_{\mathit{Top}}(9)$ versus $R_{\mathit{Top}}(10)$	&99.85\%\\\hline
$R_{\mathit{Top}}(1)$ versus $R_{\mathit{Top}}(10)$	&97.46\%\\
\hline
\end{tabular}
\label{t:stability}
\end{table}
Looking at Table \ref{t:stability}, we observe that the ranking produced using just the Top 2 programs as reference set 
has a 99.38\% of agreement with the ranking produced using just the Top 1 program. The agreement stays high (greater than 99\%) 
when adding new top programs to the reference set, one by one. Also of notice, exactly the same ranking was produced 
by the Top 5 and Top 6 rankings, and also by the Top 8 and Top 9 rankings. At the end, we compared the 
Top 1 ranking with the Top 10 ranking to observe an agreement of 97.46\%. This shows that changes in the size of the top 
reference set do not cause major changes in the final ranking. That is, these early experiments suggest that 
R-Score is a quite stable metric, relatively to the size of the top reference set. 


\section*{Conclusions}
\label{sec:conclusions}

Ranking graduate programs for the purpose of allocating research funds is a problem of significance in 
real life. A ranking of programs allows not only rewarding those that are more productive, but also 
provides a level of transparency on how public funds are spent. 

In this paper we introduced a new metric, which we refer to as R-Score for reputation-based score, based on comparing listings of publication records of whole programs in a given area of knowledge, with those of the top programs in that area. 
The idea is to use the top programs 
as referencial beacons to the other programs. In this model, programs that publish frequently in venues 
preferred by top programs fair better than programs that publish elsewhere.

To quantify the transfer of reputation, from top programs to those programs we want to compare, 
we used a Markovian model. Most important, 
transition rates in our Markov network are computed using just relative 
frequencies of publication in venues. This has two important implications: (a) access to contents of citing 
publications is not required and (b) this Markov network is simple and fast to compute. 

In our experiments, we compared an R-Score ranking of the top 25 CS programs in Brazil with a ranking provided 
by the Brazilian funding agency CAPES. For R-Score, we used the top 10 CS programs in the US, according 
to the NRC, as reference set. The results indicate very good agreement between the two rankings 
and suggest that R-Score can be useful for providing early glances into the reputation of graduate 
programs one wants to compare. 

\section*{Acknowledgements}

This work was partially sponsored
the Brazilian National Institute of Science and
Technology for the Web (grant MCT/CNPq 573871/2008-6),
and by the authors' individual grants and scholarships
from CNPq, FAPEMIG and FAPERJ.

\bibliographystyle{elsarticle-harv}
\bibliography{paper}







\end{document}